\documentclass[12pt]{article}
\usepackage{amssymb,latexsym,epsfig,cite}

\newcommand{\be}{\begin{equation}}
\newcommand{\ee}{\end{equation}}
\def\bea{\begin{eqnarray}}
\def\eea{\end{eqnarray}}
\begin{document}

\thispagestyle{empty}

\begin{center}
{\Large \bf A Primer on Resonances in Quantum Mechanics}
\end{center}

\begin{center}

Oscar Rosas-Ortiz$^1$, Nicol\'as Fern\'andez-Garc\'{\i}a$^1$ and Sara
Cruz~y~Cruz$^{1,2}$\\[2ex]
{\footnotesize $^{1}$\it Departamento de F\'{\i}sica, Cinvestav,
AP 14-740, 07000 M\'exico~DF, Mexico}\\
{\footnotesize $^{2}$\it Secci\'on de Estudios de Posgrado e
Investigaci\'on, UPIITA-IPN, Av. IPN 2508, CP 07340 M\'exico~DF,
Mexico}
\end{center}

\begin{abstract}
After a pedagogical introduction to the concept of resonance in classical
and quantum mechanics, some interesting applications are discussed. The
subject includes resonances occurring as one of the effects of radiative
reaction, the resonances involved in the refraction of electromagnetic
waves by a medium with a complex refractive index, and quantum decaying
systems described in terms of resonant states of the energy. Some useful
mathematical approaches like the Fourier transform, the complex scaling
method and the Darboux transformation are also reviewed. 
\end{abstract}

%%%%%%%%%%%%%%%%%%%%%%%%%%%%%%%%%%%%%%%%%%%%
%% MAINMATTER
%%%%%%%%%%%%%%%%%%%%%%%%%%%%%%%%%%%%%%%%%%%%

%% ----------------------------------------->Introduction
\section{Introduction}

Solutions of the Schr\"odinger equation associated to complex eigenvalues
$\epsilon = E -i\Gamma/2$ and satisfying purely outgoing conditions are
known as Gamow-Siegert functions \cite{Gam28,Bre36}. These solutions
represent a special case of scattering states for which the `capture' of
the incident wave produces delays in the scattered wave. The `time of
capture' can be connected with the lifetime of a decaying system
(resonance state) which is composed by the scatterer and the incident
wave. Then, it is usual to take ${\rm Re}(\epsilon)$ as the binding energy
of the composite while ${\rm Im}(\epsilon)$ corresponds to the inverse of
its lifetime. The Gamow-Siegert functions are not admissible as physical
solutions into the mathematical structure of Quantum Mechanics since, in
contrast with conventional scattering wave-functions, they are not finite
at $r\rightarrow \infty$. Thus, such a kind of functions is acceptable in
Quantum Mechanics only as a convenient model to solve scattering
equations. However, because of the resonance states relevance, some
approaches extend the formalism of quantum theory so that they can be
defined in a precise form \cite{Boh89,del02,del07,Civ04,Agu71,Sim72}.

The concept of resonance arises from the study of oscillating systems in
classical mechanics and extends its applications to physical theories like
electromagnetism, optics, acoustics, and quantum mechanics, among others.
In this context, resonance may be defined as the excitation of a system by
matching the frequency of an applied force to a characteristic frequency
of the system. Among the big quantity of examples of resonance in daily
life one can include the motion of a child in a swing or the tuning of a
radio or a television receiver. In the former case you must push the
swing from time to time to maintain constant the amplitude of the
oscillation. In case you want to increase the amplitude you should push
`with the natural motion' of the swing. That is, the acting of the force
you are applying on the swing should be in `resonance' with the swing
motion. On the other hand, among the extremely large number of
electromagnetic signals in space, your radio responds only to that one for
which it is tuned. In other words, the set has to be in resonance with a
specific electromagnetic wave to permit subsequent amplification to an
audible level. In this paper we present some basics of resonance
phenomenon. It is our intent to provide a strong primer introduction to
the subject rather than a complete treatment. In the next sections we
shall discuss classical models of vibrating systems giving rise to
resonance states of the energy. Then we shall review some results arising
from the Fourier transform widely used in optics and quantum mechanics.
This material will be useful in the discussions on the effects of
radiative reaction which are of great importance in the study of atomic
systems. We leave for the second part of these notes the discussion on the
resonances in quantum decaying systems and their similitudes with the
behavior of optical devices including a complex refractive index. Then
the complex scaling method arising in theories like physical chemistry is
briefly reviewed to finish with a novel application of the ancient Darboux
transformation in which the transformation function is a quantum resonant
state of the energy. At the very end of the paper some lines are included
as conclusions.

%% ----------------------------------------->Chapter 1
\section{Vibration, Waves and Resonances}

\subsection{Mechanical Models}

Ideal vibrating (or oscillating) systems undergo the same motion over and
over again. A very simple model consists of a mass $m$ at the end of a
spring which can slide back and forth without friction. The
time taken to make a complete vibration is the {\it period\/} of
oscillation while the {\it frequency\/} is the number of vibration cycles
completed by the system in unit time. The motion is governed by the
acceleration of the vibrating mass
\be
\frac{d^2x}{dt^2} = -\left(\frac{k}{m}\right) x \equiv -w_0^2 x
\label{oscila1}
\ee 
where $w_0 := \sqrt{k/m}$ is the {\it natural angular frequency\/} of the
system. In other words, a general displacement of the mass follows the
rule
\be
x= A \cos w_0 t + B \sin w_0 t
\label{oscila2}
\ee
with $A$ and $B$ two arbitrary constants. To simplify our analysis we
shall consider a particular solution by taking $A= a \cos \theta$ and $B=
-a \sin \theta$, therefore we can write
\be
x= a \cos(w_0 t + \theta).
\label{oscila3}
\ee
At $t_n = \frac{(2n+1) \pi -2\theta}{2w_0}$, $n=0,1,2\ldots$, the kinetic
energy $T= \frac12 m (dx/dt)^2$ reaches its maximum value $T_{\rm
max}=mw_0^2a^2/2$ while $x$ passes through zero. On the other hand, the
kinetic energy is zero and the displacement of the mass is maximum ($x=a$
is the {\it amplitude\/} of the oscillation)  at $t_m = \frac{m\pi -
\theta}{w_0}$, $m=0,1,2,\ldots$ This variation of $T$ is just opposite of
that of the potential energy $V=kx^2/2$. As a consequence, the total
stored energy $E$ is a constant of motion which is proportional to the
square of the amplitude (twice the amplitude means an oscillation which
has four times the energy):
\be
E=T+V= \frac12 m w_0^2 a^2. 
\label{oscila4} 
\ee
Systems exhibiting such behavior are known as {\it harmonic
oscillators\/}. There are plenty of examples: a weight on a spring, a
pendulum with small swing, acoustical devices producing sound, the
oscillations of charge flowing back and forth in an electrical circuit,
the `vibrations' of electrons in an atom producing light waves, the
electrical and magnetical components of electromagnetic waves, and so on.

%%%%%%%%%%%%%%%%%%%%
\subsubsection{Steady-state oscillations}

In actual vibrating systems there is some loss of energy due to friction
forces. In other words, the amplitude of their oscillations is a
decreasing function of time (the vibration {\it damps down\/}) and we say
the system is {\it damped\/}. This situation occurs, for example, when the
oscillator is immersed in a viscous medium like air, oil or water. In a
first approach the friction force is proportional to the velocity $F_f=
-\alpha \frac{dx}{dt}$, with $\alpha$ a {\it damping constant\/} expressed
in units of mass times frequency. Hence, external energy must be supplied
into the system to avoid the damping down of oscillations. In general,
vibrations can be driven by a repetitive force $F(t)$ acting on the
oscillator. So long as $F(t)$ is acting there is an amount of work done to
maintain the stored energy (i.e., to keep constant the amplitude). Next we
shall discuss the forced oscillator with damping for a natural frequency
$w_0$ and a damping constant $\alpha$ given.

\begin{figure}[htb]
\centering
\includegraphics[height=.18\textheight]{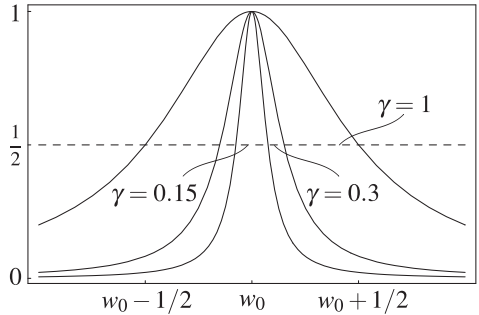}
\caption{\footnotesize The Fock-Breit-Wigner (Lorentz-Cauchy) distribution
$\omega$ for different values of the width at half maximum $\gamma$.}
\label{figura2}
\end{figure}

Let us consider an oscillating force defined as the real part of $F(t) = F
e^{iwt} \equiv F_0 e^{i(wt+\eta)}$. Our problem is to solve the equation
\be
\frac{d^2 x}{dt^2} + \gamma \, \frac{dx}{dt} + w_0^2 x = {\rm Re} \left(
\frac{F e^{iwt}}{m}\right), \qquad \gamma := \frac{\alpha}{m}.
\label{oscila5}
\ee
Here the new damping constant $\gamma$ is expressed in units of
frequency. The ansatz $x = {\rm Re} (z e^{iwt})$ reduces (\ref{oscila5}) 
to a factorizable expression of $z$, from which we get
\be
z = \frac{F \Omega}{m} , \qquad \Omega = \frac{1}{w_0^2 -w^2 +i\gamma w}.
\label{oscila6}
\ee
We realize that $z$ is proportional to the complex function $\Omega$,
depending on the driving force's frequency $w$ and parameterized by the
natural frequency $w_0$ and the damping constant $\gamma$. In polar form
$\Omega = \vert \Omega \vert e^{i\phi}$, the involved phase angle $\phi$ 
is easily calculated by noticing that $\Omega^{-1} = e^{-i\phi}/\vert
\Omega \vert = w_0^2 -w^2 + i\gamma w$, so we get
\be
\tan \phi = -\frac{\gamma w}{w_0^2 -w^2}.
\label{phase}
\ee
Let us construct a single valued phase angle $\phi$ for finite values of
$w_0$ and $\gamma$. Notice that $w<w_0$ leads to $\tan \phi<0$ while $w
\rightarrow w_0^-$ implies $\tan \phi \rightarrow -\infty$. Thereby we can
set $\phi(w=0)=0$ and $\phi(w_0) = -\pi/2$ to get $\phi \in [-\pi/2,0]$
for $w\leq w_0$. Now, since $w>w_0$ produces $\tan \phi >0$, we use
$\tan(-\phi) = -\tan \phi$ to extend the above defined domain $\phi
\in (-\pi,0]$, no matter the value of the angular frequency $w$. Bearing
these results in mind we calculate the real part of $z$ (see equation
\ref{oscila6}) to get the physical solution
\be
x= x_0 \cos(wt+\eta +\phi), \qquad x_0:= \frac{F_0 \vert \Omega \vert}{m}.
\label{oscila7}
\ee
Notice that the mass oscillation is not in phase with the driving force
but is shifted by $\phi$. Moreover, $\gamma \rightarrow 0$ produces $\phi
\rightarrow 0$, so that this phase shift is a measure of the damping.
Since the phase angle is always negative or zero ($-\pi/2 < \phi \leq 0$),
equation (\ref{oscila7}) also means that the displacement $x$ lags behind
the force $F(t)$ by an amount $\phi$. On the other hand, the amplitude
$x_0$ results from the quotient $F_0/m$ scaled up by $\vert \Omega
\vert$. Thus, such a scale factor gives us a measure of the response of
the oscillator to the action of the driving force. The total energy
(\ref{oscila4}), with $a=x_0$, is then a function of
the angular frequency:
\be
E(w)= \frac{(w_0 F_0)^2}{2m} \vert \Omega \vert^2 \equiv \frac{(w_0
F_0)^2}{2m} \left[ \frac{1}{(w_0^2 -w^2)^2 + (\gamma w)^2}\right].
\label{oscila8}
\ee
Equations (\ref{phase}) and (\ref{oscila8}) comprise the complete solution
to the problem. The last one, in particular, represents the {\it spectral
energy distribution\/} of the forced oscillator with damping we are
dealing with. It is useful, however, to simplify further under the
assumption that $\gamma <<1$. For values of $w$ closer to that of $w_0$
the energy approaches its maximum value $2m E(w_0) \approx (F_0
/\gamma)^2$ while $E(w \rightarrow +\infty)$ goes to zero as $w^{-4}$. In
other words, $E(w)$ shows rapid variations only near $w_0$. It is then
reasonable to substitute
\be
w_0^2 -w^2 = (w_0 -w)(w_0 +w) \approx (w_0 -w)2w_0
\label{resona1}
\ee
in the expressions of the energy and the phase shift to get
\be
E(w \rightarrow w_0) \approx \frac{1}{2m} \left( \frac{F_0}{\gamma}
\right)^2 \omega
(w,w_0,\gamma), \qquad \tan \phi \approx \frac{\gamma}{2(w-w_0)}
\label{resona2}
\ee
with
\be
\omega (w,w_0,\gamma):= \frac{(\gamma/2)^2}{(w_0 -w)^2 + (\gamma/2)^2}.
\label{fbw}
\ee
Equation (\ref{fbw}) describes a bell-shaped curve known as the {\it
Cauchy\/} (mathematics), {\it Lorentz\/} (statistical physics) or {\it
Fock-Breit-Wigner\/} (nuclear and particle physics) distribution. It is
centered at $w=w_0$ (the {\it location parameter\/}), with a half-width at
half-maximum equal to $\gamma/2$ (the {\it scale parameter\/}) and
amplitude (height) equal to 1. That is, the damping constant $\gamma$
defines the width of the spectral line between the half-maximum points
$w_0 -w = \pm \gamma/2$. Fig. \ref{figura2} shows the behavior of the
curve $\omega$ for different values of the damping constant
({\it spectral width\/}) $\gamma$.

%%%%%%%%%%%%%%%%%%%%%%%%%%%%%%
%%%%%%%%%%%%%%%%%%%%%%%%%

These last results show that the adding of energy to the damping
oscillator is most efficient if the vibrations are sustained at a
frequency $w = w_0$. In such a case it is said that the driving force is
in {\it resonance\/} with the oscillator and $w_0$ is called the {\it
resonance frequency\/}. Besides the above discussed spring-mass system,
the motion of a child in a swing is another simple example giving rise to
the same profile. To keep the child+swing system oscillating at constant
amplitude you must push it from time to time. To increase the amplitude
you should push `with the motion': the oscillator vibrates most strongly
when the frequency of the driving force is equal to the frequency of the 
free vibration of the system. On the other hand, if you push against the
motion, the oscillator do work on you and the vibration can be brought to
a stop. The above cases include an external force steading the
oscillations of the system. This is why equations 
(\ref{phase}--\ref{oscila8}) are known as the {\it steady-state} solutions
of the problem.

We can calculate the amount of work $W_{\rm work}$ which is done by
the driving force. This can be measured in terms of the {\it power} $P$,
which is the work done by the force per unit time:
\be
P = \frac{d}{dt}W_{\rm work} = F(t) \frac{dx}{dt} = \frac{dE}{dt} +
m\gamma \left( \frac{dx}{dt} \right)^2.
\label{power}
\ee
The {\it average\/} power $\langle P \rangle$ corresponds to the {\it
mean\/} of $P$ over many cycles. To calculate it we first notice that
$\langle dE/dt \rangle =0$. That is, the energy $E$ does not change over a
period of time much larger than the period of oscillation. Now, since the
square of any sinusoidal function has an average equal to $1/2$, the last
term in (\ref{power}) has an average which is proportional to the square
of the frequency times the amplitude of the oscillation. From
(\ref{oscila7}) we get
\be
\langle P \rangle = \frac{m \gamma w^2 x_0^2}{2}.
\label{power2}
\ee
It is then clear that the driving force does a great deal of work to
cancel the action of the friction force. In a similar form we obtain the
average of the stored energy:
\be
\langle E \rangle = \frac{mx_0^2 (w^2 + w_0^2)}{4}.
\label{power3}
\ee
Remark that the mean of $E$ does not depend on the friction but on the
angular frequency of the driving force. If $w$ is close to the resonance
frequency $w_0$, then $\langle E \rangle$ goes to the ideal oscillator's
energy (\ref{oscila4}), scaled by $(x_0/a)^2$. Moreover, the same result
is obtained no matter the magnitude of the driving force, since it does
not play any role in (\ref{power3}).

%%%%%%%%%%%%%%%%%%%%%%%%%%%%%%%%%
\subsubsection{Transient oscillations}

Suppose a situation in which the driving force is turned off at a given
time $t=t_0$. This means no work is done to sustain the oscillations so
that there is no supplied energy to preserve the motion any longer. This
system can be studied by solving (\ref{oscila5}) with $F=0$. After
introducing the ansatz $x= {\rm Re} (ze^{iwt})$ we get a quadratic
equation for $w$, the solution of which reads
\be
w_{\pm} = i \gamma/2 \pm \vartheta, \qquad \vartheta:= \sqrt{w_0^2
-(\gamma/2)^2}.
\label{trans1}
\ee
If $\gamma < w_0$ then $\vartheta \in {\bf R}$ and any of these two roots
produces the desired solution:
\be
x= {\rm Re} (z e^{iwt}) = \vert z \vert e^{-\frac{\gamma}{2} t}
\cos(\vartheta t+z_0), \quad t \geq t_0.
\label{trans2}
\ee
First, notice that the energy is not a constant of motion but
decreases in exponential form $E \propto \vert z \vert^2 e^{-\gamma t}$.
The damping constant $\gamma$ is then a measure of the lifetime of the
oscillation because at the time $\tau = 1/\gamma$, the energy is reduced
to approximately the $36\%$ ($E \rightarrow E/e$) while the amplitude goes
to the $60\%$ of its initial value ($\vert z \vert \rightarrow \vert z
\vert/\sqrt{e}$). Thus, the smaller the value of $\gamma$ the larger the
lifetime $\tau$ of the oscillation. In this way, for values of $\gamma$
such that $w_0>>\gamma/2$, the discriminant in (\ref{trans1}) becomes
$\vartheta \approx w_0$. Thereby, the system exhibits an oscillation of
frequency close to the resonance frequency $w_0$. This means that large
lifetimes are intimately connected with resonances for small values of
the damping constant.

As we can see, the resonance phenomenon is a characteristic of vibrating
systems even in absence of forces steading the oscillations. Solutions
like (\ref{trans2}) are known as {\it transient\/} oscillations because
there is no force present which can ensure their prevalentness. They are
useful to describe mechanical oscillators for which the driven force has
been turned off at the time $t=t_0$ or, more general, decaying systems
like the electric field emitted by an atom. In general, `resonance' is the
tendency of a vibrating system to oscillate at maximum amplitude under
certain frequencies $w_n$, $n=0,1,2,\dots$ At these resonance frequencies
even small driving forces produce large amplitude vibrations. The
phenomenon occurs in all type of oscillators, from mechanical and
electromagnetic systems to quantum probability waves. A resonant
oscillator can produce waves oscillating at specific frequencies. Even
more, this can be used to pick out a specific frequency from an arbitrary
vibration containing many frequencies.

%% ----------------------------------------->Chapter 1
\subsection{Fourier Optics Models}

In this section we shall review some interesting results arising from the
Fourier transform. This mathematical algorithm is useful in studying the
properties of optical devices, the effects of radiative reaction on the
motion of charged particles and the energy spectra of quantum systems as
well. Let $\{e^{ikx}\}$ be a set of plane waves orthonormalized as follows
\be
(e^{i\kappa x},e^{ikx}) = \lim_{a \rightarrow
+\infty} \int_{-a}^{a} e^{i(k-\kappa)x} dx = \lim_{a \rightarrow 
+\infty} 2 \frac{\sin(k-\kappa)a}{k-\kappa} \equiv 2\pi \delta(k-\kappa)
\label{plane1}
\ee
with $\delta(x-x_0)$ the Dirac's delta distribution. This `function'
arises in many fields of study and research as representing a sharp
impulse applied at $x_0$ to the system one is dealing with. The response
of the system is then the subject of study and is known as the {\it
impulse response\/} in electrical engineering, the {\it spread function\/}
in optics or the {\it Green's function\/} in mathematical-physics. Among
its other peculiar properties, the Dirac function is defined in such a way
that it can sift out a single ordinate in the form
\be
f(x_0) = \int_{-\infty}^{\infty} \delta (x-x_0) f(x_0) dx.
\label{delta0}
\ee
In general, a one-dimensional function $\varphi(x)$ can be expressed as
the linear combination
\be
\varphi(x) = \frac{1}{2\pi} \int_{-\infty}^{+\infty} \widetilde \varphi(k)
e^{-ikx} dk
\label{fourier1}
\ee
where the coefficient of the expansion $\widetilde \varphi(k)$ is given by
the following inner product
\be
\begin{array}{rl}
(e^{-ik x}, \varphi) & = \displaystyle \int_{-\infty}^{+\infty} \varphi(x)
e^{ikx}dx = \frac{1}{2\pi} \displaystyle \int_{-\infty}^{+\infty} \left[
\displaystyle \int_{-\infty}^{+\infty} e^{i(k-\kappa)x} dx
\right]\widetilde \varphi(\kappa) d\kappa\\[2ex]
& = \displaystyle \int_{-\infty}^{+\infty} \delta(k-\kappa)\widetilde
\varphi(\kappa) d\kappa = \widetilde \varphi(k).
\end{array}
\label{fourier2}
\ee
If (\ref{fourier1}) is interpreted as the Fourier series of $\varphi(x)$,
then the continuous index $k$ plays the role of an {\it angular spatial
frequency\/}. The coefficient $\widetilde\varphi(k)$, in turn, is called
the {\it Fourier transform\/} of $\varphi(x)$ and corresponds to the
amplitude of the {\it spatial frequency spectrum\/} of $\varphi(x)$
between $k$ and $k+dk$. It is also remarkable that functions
(\ref{fourier1}) and (\ref{fourier2}) are connected via the Parseval
formula
\be
\int_{-\infty}^{+\infty} \vert \varphi(x)\vert^2 dx =
\int_{-\infty}^{+\infty} \vert \widetilde \varphi(k) \vert^2 dk.
\label{parse1}
\ee
This expression often represents a conservation principle. For instance,
in quantum mechanics it is a conservation of probability
\cite{Foc76}. In optics, it represents the fact that all the light passing
through a diffraction aperture eventually appears distributed throughout
the diffraction pattern \cite{Hec74}. On the other hand, since $x$ and $k$
represent arbitrary (canonical conjugate) variables, if $\varphi$ were a
function of time rather than space we would replace $x$ by $t$ and then
$k$ by the angular temporal frequency $w$ to get
\be
\varphi(t) = \frac{1}{2\pi} \int_{-\infty}^{+\infty} \widetilde 
\varphi(w) e^{-iwt}dw, \qquad
\widetilde \varphi(w) = \int_{-\infty}^{+\infty} \varphi(t) e^{iwt} dt
\label{fourier3}
\ee
and
\be
\int_{-\infty}^{+\infty} \vert \varphi(t)\vert^2 dt =
\int_{-\infty}^{+\infty} \vert \widetilde \varphi (w) \vert^2 dw.
\label{parse2}
\ee
Now, let us take a time depending wave $\varphi(t)$, defined at $x=0$ by 
\be
\varphi(t) = \varphi_0 \Theta(t) e^{-\frac{\gamma}{2} t} \cos w_0 t,
\qquad \Theta(t) =\left\{
\begin{array}{ll}
1 & t>0 \\[1ex]
0 & t<0
\end{array}
\right.
\label{damped1}
\ee
Function (\ref{damped1}) is a transient oscillation as it has been defined
in the above sections. From our experience with the previous cases we know
that it is profitable to represent $\varphi(t)$ in terms of a complex
function.
In this case $\varphi = {\rm Re} (Z)$, with
\be
Z(t) = A(t) e^{-iw_0t}, \qquad A(t) = \varphi_0 \Theta(t)
\exp \left(-\frac{\gamma}{2} t \right).
\label{damped1a}
\ee
Observe that $\vert Z(t) \vert^2 = \vert A(t) \vert^2$. Then the
Parseval formula (\ref{parse2}) gives
\be
\int_{-\infty}^{+\infty} \vert A(t)\vert^2 dt =
\int_{-\infty}^{+\infty} \vert \widetilde A (w) \vert^2 dw.
\label{parse3}
\ee
Since the stored energy at the time $t$ is proportional to $\vert A(t)
\vert^2$, both integrals in equation (\ref{parse3}) give the total energy
$W$ of the wave as it is propagating throughout $x=0$. Thereby the power
involved in the oscillation as a function of time is given by $P(t) =dW/dt
\propto \vert A(t)\vert^2$. In the same manner $I_w = dW/dw \propto \vert
\widetilde A(w) \vert^2$ is the energy per unit frequency interval. Now,
from (\ref{damped1a}) we get $A(t) = Z(t) e^{iw_0 t}$, so that
\be
A(t) = \frac{1}{2\pi} \int_{-\infty}^{+\infty} \widetilde Z(w)
e^{-(w-w_0)t} dw \equiv \frac{1}{2\pi} \int_{-\infty}^{+\infty}
a(\varepsilon) e^{-i\varepsilon t} d\varepsilon
\label{parse4}
\ee
where $\varepsilon:=w-w_0$ and $a(\varepsilon) := \widetilde Z(\varepsilon
+ w_0)$. This last term is given by
\be
a(\varepsilon) = \int_{-\infty}^{+\infty} A(t) e^{i\varepsilon t} dt =
\varphi_0 \int_0^{+\infty} e^{(i\varepsilon -\frac{\gamma}{2})t} dt=
\frac{\varphi_0}{\frac{\gamma}{2} - i(w-w_0)}.
\label{parse5}
\ee
Then, up to a global constant, we have
\be
I_w = \left(
\frac{2\varphi_0}{\gamma}\right)^2  \frac{(\gamma/2)^2}{(w -w_0)^2 +
(\gamma/2)^2}.
\label{damped3}
\ee

%%%%%%%%%%%%%%%%%%%%%%%%%%%%%%
\begin{figure}[htb]
\centering
\includegraphics[height=.18\textheight]{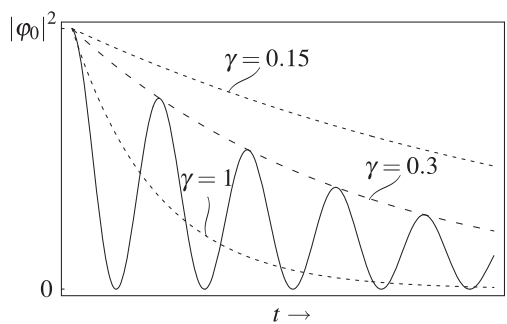}
\caption{\footnotesize The power $P(t) \propto \vert \varphi_0 \vert^2
e^{-\gamma t}$ involved with the transient oscillation (\ref{damped1}) for
the same values of $\gamma$ as those given in Figure \ref{figura2}. Remark
that the area under the dotted curves is larger for smaller values of the
line breadth $\gamma$.}
\label{figura3}
\end{figure}
%%%%%%%%%%%%%%%%%%%%%%%%%

\noindent 
That is, the Fock-Breit-Wigner (FWB) function (\ref{damped3}) is the
spectral energy distribution of the transient oscillation (\ref{damped1}).  
In the previous section we learned that the inverse of the damping
constant $\gamma$ measures the oscillation lifetime. The same rule holds
for the power $\vert A(t) \vert^2 =\vert \varphi_0 \vert^2 e^{-\gamma t}$,
as it is shown in Fig. \ref{figura3}. Let us investigate the extremal case
of infinite lifetimes. Thus we calculate $I_w$ in the limit $\gamma
\rightarrow 0$. The result reads
\be
I_w \rightarrow 2\pi \left(\frac{\varphi_0^2}{\gamma}\right)
\delta(w-w_0),
\qquad \gamma \rightarrow 0
\label{limit1}
\ee
where we have used
\be
\delta(x) = \frac{1}{\pi} \lim_{\epsilon \rightarrow 0}
\left(\frac{\epsilon}{x^2 + \epsilon^2} \right).
\label{delta}
\ee
Infinite lifetimes ($1/\gamma =\tau \rightarrow +\infty$) correspond to
spectral energy distributions of infinitesimal width ($\gamma \rightarrow
0$) and very high height ($\varphi_0^2/\gamma$). As a consequence, the
transient oscillation (\ref{damped1}) has a definite frequency ($w
\rightarrow w_0$) along the time. The same conclusion is obtained for
steady-state oscillations by taking $F_0^2 = 8m \varphi_0^2$ in equations
(\ref{resona2}--\ref{fbw}). In the next sections we shall see that this
lifetime$\leftrightarrow$width relationship links bound and decaying
energy states in quantum mechanics.

As a very simple application of the above results let the plane wave
(\ref{damped1}) be the electric field emitted by an atom. Then $W$
corresponds to the total energy radiated per unit area perpendicular to
the direction of propagation. Equation (\ref{damped3}) in turn, relates in
quantitative way the behavior of the power radiated as a function of the
time to the frequency spectrum of the energy radiated. To give a more
involved example let us consider a nonrelativistic charged particle of
mass $m_e$ and charge $q_e$, acted on by an external force $\vec F$. The
particle emits radiation since it is accelerated. To account for this
radiative energy loss and its effect on the motion of the particle it is
necessary to add a {\it radiative reaction force\/} $\vec F_{\rm rad}$ in
the equation of motion to get
\be
\vec F +\vec F_{\rm rad} \equiv \vec F +
\left( \frac{2q_e^2}{3c^3} \right) \frac{d^3}{dt^3} 
\vec r = m_e\frac{d^2}{dt^2}\vec r
\label{motion}
\ee
with $c$ the speed of light. This last expression is known as the {\it
Abraham-Lorentz} equation and is useful only in the domain where the
reactive term is a small correction, since the third order derivative term
does not fulfill the requirements for a dynamical equation (see e.g.,
reference \cite{Jac99}, Ch 16). Bearing this condition in mind let us
investigate the effect of an external force of the form $\vec F = -m_e
w_0^2 \vec r$. The Abraham-Lorentz equation is written
\be
\left( \frac{2q_e^2}{3m_e c^3} \right) \frac{d^3}{dt^3} \vec r =
\frac{d^2}{dt^2}\vec r +w_0^2 \vec r.
\label{motion2}
\ee
For small values of the third order term one has $\frac{d^2}{dt^2} \vec r
\approx -w_0^2 \vec r$. That is, the particle oscillates like a mass at
the end of a spring with frequency $w_0$. Hence $\frac{d^3}{dt^3} \vec r
\approx -w_0^2 \frac{d}{dt}\vec r$, so that the problem is reduced to the
transient equation
\be
\frac{d^2}{dt^2} \vec r +\gamma \frac{d}{dt}\vec r + w_0^2 \vec r=0,
\qquad \gamma = \frac{2q_e^2w_0^2}{3m_ec^3},
\label{para}
\ee
the solution of which has the form (\ref{damped1}-\ref{damped1a}) with
$\varphi(t)$ replaced by $\vec r(t)$ and $\varphi_0$ by a constant vector
$\vec r_0$. To get an idea of the order of our approach let us evaluate
the quotient $\gamma/w_0^2$. A simple calculation gives the following
constant
\be
\frac{\gamma}{w_0^2} \approx 0.624 \times 10^{-23} s.
\label{para2}
\ee
Thus the condition $\gamma <<1 s^{-1}$ defines the appropriate values of
the frequency $w_0$. For instance, let $\gamma$ take the value $10^{-3}
s^{-1}$. The appropriate value of $w_0$ is then of the order of an
infrared frequency $w_0 \sim 10^{10} s^{-1}$. But if $\gamma \sim 10^{-11}
s^{-1}$ then $w_0 \sim 10^6 s^{-1}$, that is, the particle will oscillate
at a radio wave frequency.

Under the limits of our approach the radiative reaction force $\vec F_{\rm
rad}$ plays the role of a friction force which damps the oscillations of
the electric field. The {\it resonant line shape\/} defined by the
Fock-Breit-Wigner function (\ref{damped3}) is broadened and shifted in
frequency due to the reactive effects of radiation. That is, because the
decaying of the power radiated $P(t) \propto \vert \varphi_0 \vert^2
e^{-\gamma t}$, the emitted radiation corresponds to a pulse (wave train)
with effective length $\lambda \approx c/\gamma$ and covering an interval
of frequencies equal to $\gamma$ rather than being monochromatic. The
infinitesimal finiteness of the width in the spectral energy distribution
(\ref{limit1})  is then justified by the `radiation friction'. In the
language of radiation the damping constant $\gamma$ is known as the {\it
line breadth\/}.

Finally, it is well known that the effects of radiative reaction are of
great importance in the detailed behavior of atomic systems. It is then
remarkable that the simple plausibility arguments discussed above led to
the qualitative features derived from the formalism of quantum
electrodynamics. By proceeding in a similar manner, it is also possible to
verify that the scattering and absorption of radiation by an oscillator
are also described in terms of FBW-like distributions appearing in the
scattering cross section. The reader is invited to review the approach in
classical references like \cite{Jac99,Land87}.

%% ----------------------------------------->Section

\subsection{Fock's Energy Distribution Model}

\noindent
Let $\{ \phi_E(x) \}$ be a set of eigenfunctions belonging to energy
eigenvalues $E$ in the continuous spectrum of a given one-dimensional
Hamiltonian $H$. The vectors are orthonormalized as follows
\be
\int_{-\infty}^{\infty} \overline{\phi}_E(x) \phi_{E'}(x) dx = \delta (E'
-E).
\label{basis1}
\ee
Notice we have taken for granted that the continuous spectrum is not
degenerated, otherwise equation (\ref{basis1}) requires some
modifications. Let us assume that a wave function $\psi_0(x)$ can be
expanded in a series of these functions, that is:
\be
\psi_0 (x) = \int_{-\infty}^{\infty} C(E) \phi_E(x) dE, \qquad 
C(E) = \int_{-\infty}^{+\infty} \overline{\phi}_E(x) \psi_0(x) dx.
\label{basis2}
\ee
The inner product of $\psi_0$ with itself leads to the Parseval relation
\be
W:=(\psi_0,\psi_0) = \int_{-\infty}^{+\infty} \vert {\psi}_0(x) \vert^2 
dx = \int_{-\infty}^{+\infty} \vert C(E) \vert^2 dE.
\label{basis2a}
\ee
In the previous sections we learned that $W$ allows the definition of
the energy distribution $\omega(E)$. In this case we have
\be
\omega(E):= \frac{dW}{dE} = \vert C(E) \vert^2.
\label{basis2b}
\ee
At an arbitrary time $t>0$ the state of the system reads
\be
\psi_t (x) = \int_{-\infty}^{\infty} C(E) \phi_E(x) e^{-iEt/\hbar} dE.
\label{basis3}
\ee
The {\it transition amplitude\/} $T(t \geq 0)$ from the state $\psi_0$
into $\psi_t$ is given by the inner product
\be
T(t \geq 0) \equiv (\psi_0, \psi_t) = \int_{-\infty}^{+\infty}
\overline{\psi}_0 (x) \psi_t(x) dx = \int_{-\infty}^{+\infty} \omega(E)
e^{-iEt/\hbar} dE.
\label{basis4}
\ee
Function $T$ rules the transition probability $\vert T (t) \vert^2$ from
$\psi_0(x)$ to $\psi_t (x)$ by relating the wave function at two different
times: $t_0$ and $t \geq t_0$. It is known as the {\it propagator\/} in
quantum mechanics and can be identified as a (spatial) Green's function
for the time-dependent Schr\"odinger equation (see next section). From
(\ref{basis4}), it is clear that $T$ can be investigated in terms of
spatial coordinates $x$ or as a function of the energy distribution. Next,
following the Fock's arguments \cite{Foc76}, we shall analyze the
transition probability for a decaying system by assuming that $\omega(E)$
is given. Let $\omega(E)$ be the Fock-Breit-Wigner distribution
\be
\omega(a) = \frac{1}{\pi} \left[\frac{(\Gamma/2)^2}{a^2 + 
(\Gamma/2)^2} \right] =
\frac{1}{\pi} \left[\frac{(\Gamma/2)^2}{(a+i\Gamma/2)(a-i\Gamma/2)} 
\right], \quad  a: =E-E_0. 
\label{fock1}
\ee
Then equation (\ref{basis4}) reads
\be
T(t\geq 0) = \frac{1}{\pi} \left( \frac{\Gamma}{2} \right)^2 e^{-iE_0
t/\hbar} \int_{-\infty}^{+\infty} \frac{e^{-iat/\hbar}}{(a+i\Gamma/2)
(a-i\Gamma/2)}.
\label{basis5}
\ee

%%%%%%%%%%%%%%%%%%%%%%%%%%%%%%
\begin{figure}[htb]
\centering
\includegraphics[height=.18\textheight]{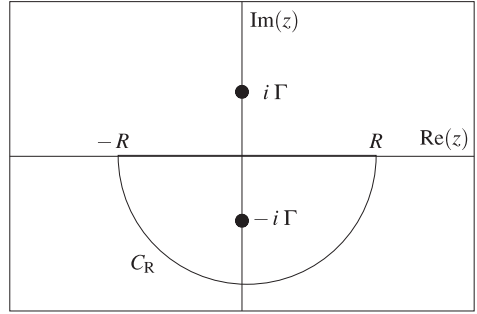}
\caption{\footnotesize Contour of integration in the complex $E$-plane.}
\label{contour}
\end{figure}
%%%%%%%%%%%%%%%%%%%%%%%%%

\noindent
We start with the observation that function (\ref{fock1}) has two isolated
singularities at $a=\pm i\Gamma/2$ (none of them lies on the real axis!).
When $R>0$, the point $-i\Gamma/2$ lies in the interior of the
semicircular region which is depicted in Fig. \ref{contour}. At this
stage,
it is convenient to introduce the function
\be
f(z) = \frac{g(z)}{z+i\Gamma/2}, \qquad g(z) =
\frac{e^{-izt/\hbar}}{z-i\Gamma/2}.
\label{fock2}
\ee
Integrating $f(z)$ counterclockwise around the boundary of the
semicircular region the {\it Cauchy integral formula\/} (see e.g.,
\cite{Bro03}) is written
\be
\int_{R}^{-R} f(a) da + \int_{C_R} f(z) dz = 2\pi i
g (-i \Gamma/2).
\label{fock3}
\ee
This last expression is valid for all values of $R>0$. Since the value of
the integral on the right in (\ref{fock3}) tends to 0 as $R \rightarrow
+\infty$, we finally arrive at the desired result:
\be
\int_{-\infty}^{+\infty} f(a) da = \frac{2\pi}{\Gamma} \exp \left(
-\frac{\Gamma t}{2 \,\hbar} \right).
\label{fock4}
\ee
In this way the transition rate $T(t)$ acquires the form of a transient
oscillation
\be
T(t \geq 0) = \frac{\Gamma}{2} \exp (-i\epsilon t/\hbar), \qquad \epsilon
:= E_0 -i\Gamma/2.
\label{fock5}
\ee
It is common to find $T(t)$ as free of the factor $\Gamma/2$ (see e.g.,
\cite{Foc76}, pp 159). This factor arises here because (\ref{fock1}) has
been written to be consistent with the previous expressions of a FWB
distribution. The factor is easily removed if we take $\Gamma/2$ rather
than $(\Gamma/2)^2$ in the numerator of $\omega(a)$. Now, we know that
transient oscillations involve lifetimes and the present case is not an
exception. Equation (\ref{fock5}) means that the transition from
$\psi_0(x)$ to $\psi_t(x)$ is an exponential decreasing function of the
time. Since this rate of change is symmetrical, it also gives information
about the rate of decaying of the initial wave. Thus, the probability that
the system has not yet decayed at time $t$ is given by
\be
\vert T(t) \vert^2 = \left( \frac{\Gamma}{2} \right)^2 \exp(-\Gamma
t/\hbar), \quad t\geq 0.
\label{fock6}
\ee
We note that the state of the undecayed system $\psi_0$ does not change
but decays suddenly. That is, the time $t$ in (\ref{fock6}) is counted off
starting from the latest instant when the system has not decayed. The
above description was established by Fock in his famous book on quantum
mechanics \cite{Foc76} and this is why functions like (\ref{fock1}) bears
his name. It is remarkable that the first Russian edition is
dated on August 1931.

Finally, remark we have written (\ref{fock5}) in terms of the complex
number $\epsilon = E_0 -i\Gamma/2$. The reason is not merely aesthetic
because, up to a constant factor, $T(t \geq 0)$ is the Fourier
transform\footnote{The Fourier transform in this case corresponds to the
equations (\ref{fourier3}), with the energy $E$ and the angular frequency
$w$ related by the Einstein's expression $E=\hbar w$.} of the expansion
coefficient $C(E)$:
\be
C(E) := \frac{\Gamma/2}{\sqrt \pi (E-E_0+i\Gamma/2)} 
= \frac{\Gamma/2}{\sqrt \pi (E-\epsilon)} 
\label{fock7}
\ee
where we have used (\ref{fock1}). The relevance of this result will be
clarified in the sequel.

%%%%%%%%%%%%%%%%%%%%%%%%%%%%%%%%%%%%%%%%%%%%%
\section{Quanta, Tunneling and Resonances}

Let us consider the motion of a particle of mass $m$ constrained to move
on the straight-line in a given potential $U(x)$. Its
{\it time-dependent Schr\"odinger equation\/} is
\be
H\psi(x,t) := \left[- \frac{\hbar^2}{2m} \frac{\partial^2}{\partial x^2}
+U(x) \right] \psi(x,t) = i\,\hbar \frac{\partial}{\partial t} \psi(x,t).
\label{schro1}
\ee
Let us assume the wave-function $\psi(x,t)$ is separable, that is
$\psi(x,t) = \varphi(x) \theta(t)$. A simple calculation leads to $\theta
(t) = \exp (-i Et/\hbar)$, with $E$ a constant, and $\varphi(x)$ a
function fulfilling
\be
\left[- \frac{\hbar^2}{2m} \frac{\partial^2}{\partial x^2}
+U(x) \right] \varphi(x) \equiv H \varphi(x) = E \varphi(x).
\label{schro2}
\ee
This time-independent Schr\"odinger equation (plainly the
Schr\"odinger equation) defines a set of {\it eigenvalues\/} $E$ and {\it
eigenfunctions\/} of the Hamiltonian operator $H$ which, in turn,
represents the {\it observable\/} of the energy. Now, to get some
intuition about the separableness of the wave-function let us take the
Fourier transform of its temporal term
\be
\widetilde \theta (E) = \lim_{a\rightarrow +\infty} \int_{-a}^{a}
e^{i(E-E')t/\hbar} dt = 2\pi \delta (E-E')
\label{sta1}
\ee
where we have used (\ref{plane1}). This last result means that the energy
distribution is of infinitesimal width and very high height. In other
words, the system has a definite energy $E=E'$ along the time. Systems
exhibiting this kind of behavior are known as {\it stationary\/} and it
is said that they are {\it conservative\/}. Since the Hamiltonian operator
$H$ does not depend on $t$ and because for any analytic function $f$ of
$H$ one has
\be
f(H) \varphi(x) = f(E) \varphi(x)
\label{sta2}
\ee
our separability ansatz $\psi \rightarrow \varphi \theta$ can be written
\be
\psi(x,t) = \exp\left( -\frac{i}{\hbar} Ht \right) \varphi(x) =
e^{-iEt/\hbar} \varphi(x).
\label{sta3}
\ee
According with the Born's interpretation, the wave function $\varphi(x)$
defines the probability density $\rho(x) =\vert \varphi (x) \vert^2$
of finding the quantum particle between $x$ and $x+dx$. Thereby, the sum
of all probabilities (i.e., the probability of finding the particle
anywhere in the straight-line at all) is unity:
\be
\int_{-\infty}^{+\infty} \rho(x) dx = \int_{-\infty}^{+\infty}
\overline{\varphi}(x) \varphi(x) dx \equiv \int_{-\infty}^{+\infty}
\vert \varphi(x) \vert^2 dx =1.
\label{norma}
\ee
The above equation represents the {\it normalization condition\/}
fulfilled by the solutions $\varphi(x)$ to be physically acceptable.
Hence, they are elements of a vector space ${\cal H}$ consisting of {\it
square-integrable\/} functions and denoted as ${\cal H} = L^2 ({\bf R},
\mu)$, with $\mu$ the Lebesgue measure (for simplicity in notation we
shall omit $\mu$ by writing $L^2({\bf R})$). As an example, ${\cal H}$
can be the space spanned by the Hermite polynomials $H_n(x)$, weighted by
the factor $\mu(x) =e^{-x^2/2}$ and defined as follows
\be
H_n(x) = (-1)^n e^{x^2/2} \frac{d^n}{dx^n} e^{-x^2/2}.
\label{hermite}
\ee
In quantum mechanics, observables are represented by the so-called
Hermitian operators in the Hilbert space ${\cal H}$. A differential
operator $A$ defined on $L^2({\bf R})$ is said to be {\it Hermitian\/} if,
whenever $Af$ and $Ag$ are defined for $f, g \in L^2({\bf R})$ and belong
to $L^2({\bf R})$, then
\be
(Af,g)=(f,Ag) = \int_{-\infty}^{+\infty} \overline{f}(x) A g(x) \mu(x) dx.
\label{hermit1}
\ee
In particular, if $f=g$ the above definition means that the action of the
Hermitian operator $A$ on $f \in L^2({\bf R})$ is symmetrical. If $Af(x) =
\alpha f(x)$, we have $(Af,f)= \overline{\alpha}$ and $(f,Af)=\alpha$, so
that $\overline{\alpha} =\alpha \Rightarrow \alpha \in {\bf R}$. In other
words, the eigenvalues of a Hermitian operator acting on $L^2({\bf R})$
are real numbers. It is important to stress, however, that this rule is
not true in the opposite direction. In general, as we are going to see in
the next sections, there is a wide family of operators $A_{\lambda}$
sharing the same set of real eigenvalues $\{ \alpha \}$. (The family
includes some non-Hermitian operators!) Moreover, notice that the rule is
not necessarily true if $f(x) \notin L^2({\bf R})$. In general, the set of
solutions of $Af(x) =\alpha f(x)$ is wider than $L^2({\bf R})$. That is,
the complete set of mathematical solutions $\varphi(x)$ embraces functions
such that its absolute value $\vert \varphi(x) \vert$ diverges even for
real eigenvalues. A plain example is given by the solutions representing
scattering states because they do not fulfill (\ref{norma}). In such a
case one introduces another kind of normalization like that defined in
(\ref{basis1}), with a similar notion of the Born's probability as we have
seen in the previous section. Normalization (\ref{norma}) is then a very
restrictive condition picking out the appropriate physical solutions among
the mathematical ones. This is why the Schr\"odinger equation
(\ref{schro1}) is ``physically solvable'' for a very narrow set of
potentials.

If $E$ is real, the time-dependent factor in (\ref{sta3}) is purely
oscillatory (a phase) and the time displacement $\psi(x,t)$ gives the same
`prediction' (probability density) as $\varphi(x)$. As a result, both of
these vectors lead to the same expectation values of the
involved observables
\be
\langle A \rangle: = \int_{-\infty}^{+\infty} \overline{\psi}(x,t) A
\psi(x,t) dx = \int_{-\infty}^{+\infty} \overline{\varphi}(x) A
\varphi(x) = \alpha.
\label{expec}
\ee
In particular, $\langle H \rangle = E$ shows that the eigenvalue $E$ is
also the expectation value of the energy. Notice that stationary states
are states of well-defined energy, $E$ being the definite value of its
energy and not only its expectation value (see equation \ref{sta1}). That
is, any determination of the energy of the particle always yields the
particular value $E$. Again, as an example, let us consider a scattering
state. Far away from the influence of the scatterer, it is represented by
a plane wave like
\be
\psi(x,t) = \exp \left(-\frac{i}{\hbar}Et \right) \exp
\left(-\frac{i}{\hbar} xp\right)
\label{scat1}
\ee
where $p$ is the linear momentum of the particle. Function (\ref{scat1})
represents a state having a definite value for its energy. However, there
is no certainty neither on the position of the particle nor in the transit
time of the particle at a given position. In general, the energy
distribution will not be a continuous function. It could include a set of
isolated points (discrete energy levels) and/or continuous portions
showing a set of very narrow and high peaks (resonance levels). The former
correspond to infinite lifetime states (the observed discrete energy
levels of atoms are good examples) while the lifetime of the second ones
will depend on the involved interactions.

%%%%%%%%%%%%%%%%%%%%%%%%%%%%%%%%%%%%%%%%%
\subsection{Quasi-stationary States and Optical Potentials}

It is also possible to define a {\it probability current density\/} $j$
\be
j = \frac{\hbar}{2mi} \left[\overline{\psi} \left(\frac{d}{dx} \psi
\right) - \left(\frac{d}{dx} \overline{\psi} \right) \psi \right]
\label{current}
\ee
which, together with the probability density $\rho = \vert \psi \vert^2$
satisfies a continuity equation
\be
\frac{d\rho}{dt} + \frac{dj}{dx} =0
\label{continuity}
\ee
exactly as in the case of conservation of charge in electrodynamics.
Observe that stationary states fulfill $d\rho /dt =0$, so
that $\rho \neq \rho(t)$ and $j=0$. What about decaying systems for which
the transition amplitude $T(t\geq 0)$ involves a complex number $\epsilon
= E_0 -i\Gamma/2$ like that found in equation (\ref{fock5})? Let us assume
a complex eigenvalue of the energy $H\varphi_{\epsilon} =
\epsilon \varphi_{\epsilon}$ is admissible in (\ref{sta3}). Then we have
$\rho(x,t) = \rho_{\epsilon}(x) e^{-\Gamma t/\hbar}$ and $j\neq 0$. That
is, complex energies are included at the cost of adding a non-trivial
value of the probability current density $j$. A conventional way to solve
this `problem' is to consider a complex potential $U=U_R +iU_I$. Then 
equation (\ref{continuity}) acquires the form
\be
\frac{d\rho}{dt} + \frac{dj}{dx} = \frac{2}{\hbar} U_I(x) \rho(x),
\label{continuity2}
\ee
the integration of which can be identified with the variation of the
number of particles
\be
\frac{d}{dt}N= \frac{2}{\hbar} \int_{-\infty}^{+\infty} U_I(x) \rho(x) dx.
\label{numer}
\ee
Let $U_I(x) = U_0$ be a constant. If $U_0>0$, there is an increment of the
number of particles ($dN/dt >0$) and viceversa, $U_0<0$ leads to a
decreasing number of particles ($dN/dt <0$). In the former case the
imaginary part of the potential works like a source of particles while the
second one shows $U_I$ as a sink of particles. The introduction of this
potential into the Schr\"odinger equation gives
\be
\left[- \frac{\hbar^2}{2m} \frac{\partial^2}{\partial x^2}
+U_R(x) +iU_0\right] \varphi_{\epsilon}(x) = \left( E_0 -i\frac{\Gamma}{2}
\right) \varphi_{\epsilon}(x).
\label{schro3}
\ee
Then, the identification $U_0 = -\Gamma/2$ reduces the solving of this
last equation to a stationary problem. The exponential decreasing
probability density $\rho_{\epsilon}(x) e^{-\Gamma t/\hbar}$ is then
justified by the presence of a sink-like potential $U_I(x)= U_0 <0$.
However, this solution requires the introduction of a non-Hermitian
Hamiltonian because the involved potential is complex. Although such a
Hamiltonian is not an observable in the sense defined in the above
section, notice that $U_0 =-i\Gamma/2$ is a kind of damping constant. The
lifetime of the probability $\rho(x,t)$ is defined by the inverse of
$\Gamma$ and, according with the derivations of the previous section, the
energy distribution shows a bell-shaped peak at $E=E_0$. In other words,
the complex eigenvalue $\epsilon =E_0 -i\Gamma/2$ is a pole of $\omega(E)$
and represents a resonance of the system. The modeling of decay by the use
of complex potentials with constant imaginary part is know as the {\it
optical model\/} in nuclear physics. The reason for such a name is
clarified in the next section.

%%%%%%%%%%%%%%%%%%%%%%%%%%%%
\subsubsection{Complex Refractive index in Optics}

It is well known that the spectrum of {\it electromagnetic energy\/}
includes radio waves, infrared radiation, the visible spectrum of colors
red through violet, ultraviolet radiation, $x$-rays and gamma radiation.
All of them are different forms of light and are usually described as {\it
electromagnetic waves\/}. The physical theory treating the {\it
propagation\/} of light is due mainly to the work of James Clerk Maxwell
(1831-1879). The interaction of light and matter or the absorption and
emission of light, on the other hand, is described by the quantum theory.
Hence, a consistent theoretical explanation of all optical phenomena is
furnished jointly by Maxwell's electromagnetic theory and the quantum
theory. In particular, the speed of light $c=299,792,456.2 \pm 1.1 m/s$ is
a part of the {\it wave equation} fulfilled by the electric field $\vec E$
and the magnetic field $\vec H$:
\be
\nabla^2 \vec A = \frac{1}{c^2} \frac{\partial^2 \vec A}{\partial t^2},
\qquad \vec A = \vec E, \vec H.
\label{max}
\ee
The above expression arises from the Maxwell's equations in empty space
with $c=(\mu_0 \epsilon_0^*)^{-1/2}$. The constant $\mu_0$ is known as the
{\it permeability of the vacuum\/} and the constant $\epsilon_0^*$ is
called the {\it permitivity of the vacuum\/} \cite{Jac99}. In isotropic
nonconducting media these constants are replaced by the corresponding
constants for the medium, namely $\mu$ and $\epsilon^*$. Consequently, the
speed of propagation $\nu$ of the electromagnetic fields in a medium is
given by $\nu = (\mu \epsilon^*)^{-1/2}$. The {\it index of refraction\/}
$n$ is defined as the ratio of the speed of light in vacuum to its speed
in the medium: $n=c/\nu$. Most transparent optical media are nonmagnetic
so that $\mu/\mu_0 =1$, in which case the index of refraction should be
equal to the square root of the relative permitivity $n =(\epsilon^*/
\epsilon^*_0)^{1/2}\equiv K^{1/2}$. In a nonconducting, isotropic medium,
the electrons are permanently bound to the atoms comprising the medium and
there is no preferential direction. This is what is meant by a simple
isotropic dielectric such a glass \cite{Fow75}. Now, consider a
(one-dimensional) plane harmonic wave incident upon a plane boundary
separating two different optical media. In agreement with the phenomena of
reflection and refraction of light ruled by the Huygen's principle, there
will be a reflected wave and a transmitted wave (see e.g. \cite{Hec74}).
Let the first medium be the empty space and the second one having a
complex index of refraction
\be
{\cal N} = n+i n_I.
\label{refra1}
\ee
Then, the wavenumber of the refracted wave is complex:
\be
{\cal K} = k + i \alpha
\label{refra2}
\ee
and we have
\[
\begin{array}{ll}
\vec E = \vec E_0 \, e^{i(k_0 x - wt)} & \mbox{\rm (incident wave)}\\[1ex]
\vec E' = \vec E_0' \, e^{i(k_0' x - wt)} & \mbox{\rm (reflected
wave)}\\[1ex]
\vec E'' = \vec E_0'' \, e^{i({\cal K} x - wt)} = \vec E_0'' \, e^{-\alpha
x} e^{i(k x -wt)} & \mbox{\rm (refracted wave)}
\end{array}
\]
If $\alpha >0$ the factor $e^{-\alpha x}$ indicates that the amplitude of
the wave decreases exponentially with the distance. That is, the energy of
the wave is absorbed by the medium and varies with distance as $e^{-2
\alpha x}$. Hence $2 \alpha$ is the {\it coefficient of absorption\/} of
the medium. The imaginary part $n_I$ of ${\cal N}$, in turn, is known as
the {\it extinction index\/}. In general, it can be shown that the
corresponding polarization is similar to the amplitude formula for a
driven harmonic oscillator \cite{Fow75}. Thus, an optical resonance
phenomenon will occur for light frequencies in the neighborhood of the
resonance frequency $w_0 = (K/m)^{1/2}$. The relevant aspect of these
results is that a complex index of refraction ${\cal N}$ leads to a
complex wavenumber ${\cal K}$. That is, the properties of the medium (in
this case an absorbing medium) induce an specific behavior of the
electromagnetic waves (in this case, the exponentially decreasing of the
amplitude). This is the reason why the complex potential discussed in the
previous section is named the `optical potential'.

%%%%%%%%%%%%%%%%%%%%%%%%%%%%%%%%%%%%%%%%%%%%%%%%%%%%%%%
\subsection{Quantum Tunneling and Resonances}

In quantum mechanics the complex energies were studied for the first time
in a paper by Gamow concerning the alpha decay (1928) \cite{Gam28}. In a
simple picture, a given nucleus is composed in part by alpha particles
(${}_2^4 He$ nuclei) which interact with the rest of the nucleus via an
attractive well (obeying the presence of nuclear forces) plus a potential
barrier (due, in part, to repulsive electrostatic forces). The former
interaction constrains the particles to be bounded while the second holds
them inside the nucleus. The alpha particles have a small (non--zero)
probability of tunneling to the other side of the barrier instead of
remaining confined to the interior of the well. Outside the potential
region, they have a finite lifetime. Thus, alpha particles in a nucleus
should be represented by {\it quasi--stationary\/} states. For such
states, if at time $t=0$ the probability of finding the particle inside
the well is unity, in subsequent moments the probability will be a slowly
decreasing function of time (see e.g. Sections 7 and 8 of reference
\cite{Foc76}). In his paper of 1928, Gamow studied the escape of alpha
particles from the nucleus via the tunnel effect. In order to describe
eigenfunctions with exponentially decaying time evolution, Gamow
introduced energy eigenfunctions $\psi_G$ belonging to complex eigenvalues
$Z_G = E_G - i\, \Gamma_G$, $\Gamma_G>0$. The real part of the eigenvalue
was identified with the energy of the system and the imaginary part was
associated with the inverse of the lifetime. Such `decaying states' were
the first application of quantum theory to nuclear physics.

Three years later, in 1931, Fock showed that the law of decay of a
quasi--stationary state depends only on the energy distribution function
$\omega(E)$ which, in turn, is meromorphic \cite{Foc76}. According to
Fock, the analytical expression of $\omega(E)$ is rather simple and has
only two poles $E= E_0 \pm i \, \Gamma$, $\Gamma >0$ (see our equation
(\ref{fock1}) and equation (8.13) of \cite{Foc76}). A close result was
derived by Breit and Wigner in 1936. They studied the cross section of
slow neutrons and found that the related energy distribution reaches its
maximum at $E_R$ with a half--maximum width $\Gamma_R$. A resonance is
supposed to take place at $E_R$ and to have ``half--value breath''
$\Gamma_R$ \cite{Bre36}. It was in 1939 that Siegert introduced the
concept of a purely outgoing wave belonging to the complex eigenvalue
$\epsilon = E-i\Gamma/2$ as an appropriate tool in the studying of
resonances \cite{Sie39}. This complex eigenvalue also corresponds to a
first--order pole of the $S$ matrix \cite{Hei47} (for more details see
e.g. \cite{del03}). However, as the Hamiltonian is a Hermitian operator,
then (in the Hilbert space ${\cal H}$) there can be no eigenstate having a
strict complex exponential dependence on time. In other words, decaying
states are an approximation within the conventional quantum mechanics
framework. This fact is usually taken to motivate the study of the rigged
(equipped) Hilbert space $\overline {\cal H}$ \cite{Boh89,del02,del07}
(For a recent review see \cite{Civ04}). The mathematical structure of
$\overline {\cal H}$ lies on the nuclear spectral theorem introduced by
Dirac in a heuristic form \cite{Dir58} and studied in formal rigor by
Maurin \cite{Mau68} and Gelfand and Vilenkin \cite{Gel68}.

\begin{figure}[htb]
\centering
\includegraphics[height=.18\textheight]{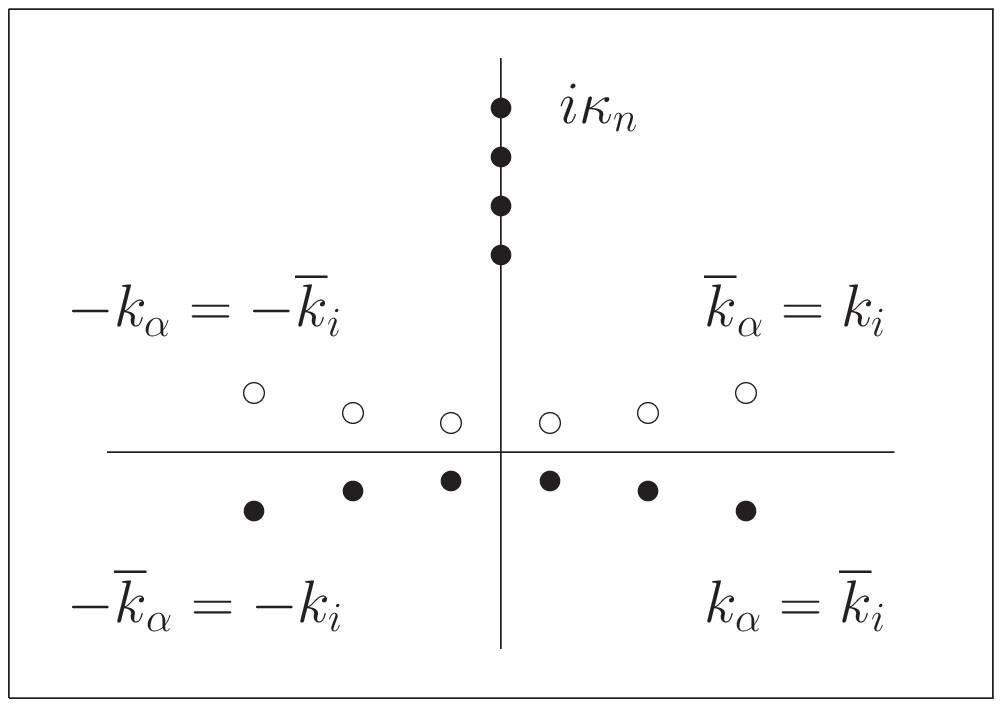}\hskip10mm
\includegraphics[height=.18\textheight]{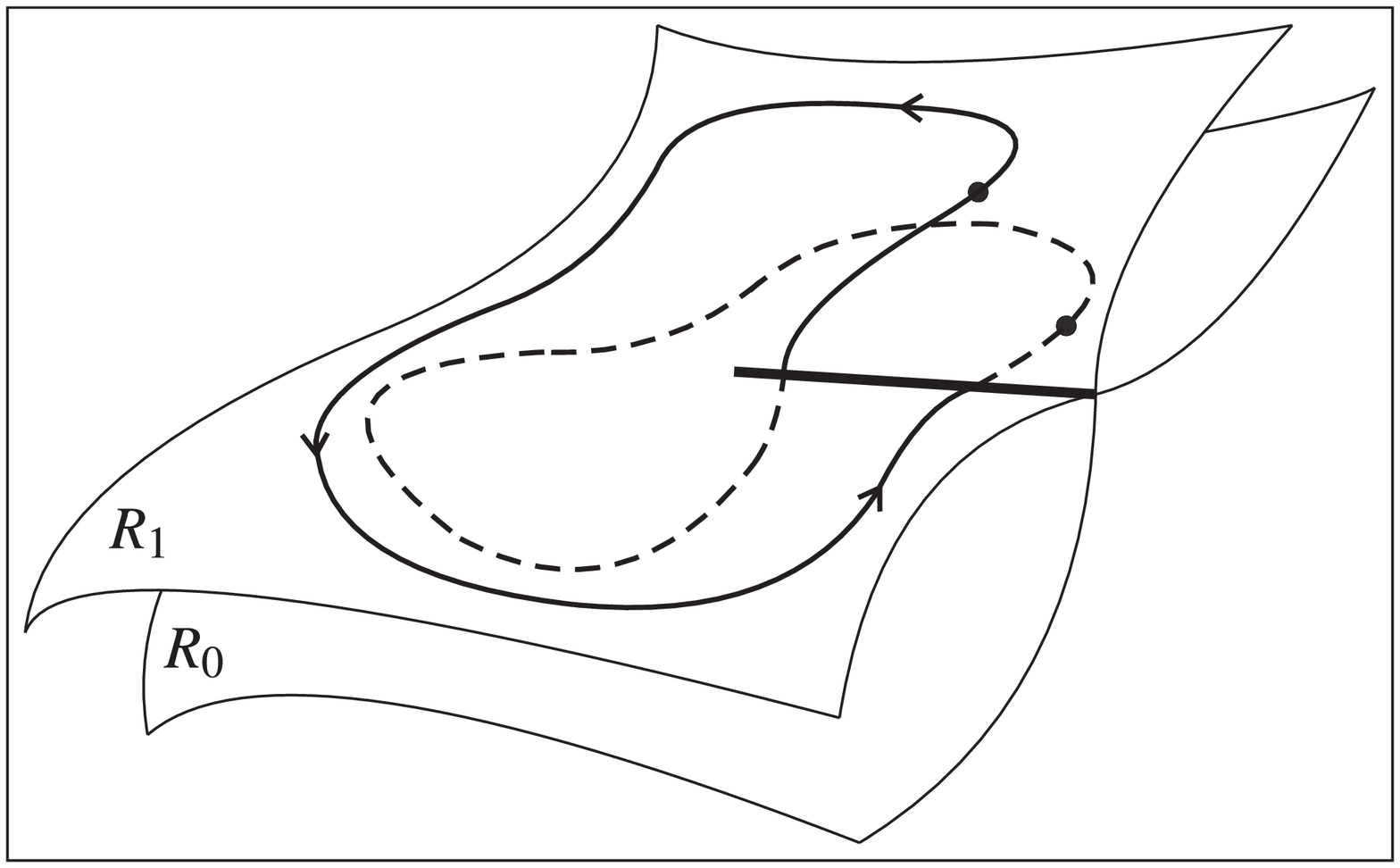}
\caption{\footnotesize {\bf Left:} Schematic representation of the poles
(disks) and the zeros (circles) of the transmission amplitude $S(k)$ in
the complex $k$-plane. Bounded energies correspond to poles located on the
positive imaginary axis {\bf Right:} The two-sheet Riemann surface
$\sqrt{\epsilon}=k$. The lower edge of the slit in $R_0$ is joined to the
upper edge of the slit in $R_1$, and the lower edge of the slit in $R_1$
is joined to the upper edge of the slit in $R_0$. The picture is based on
the description given by J.M. Brown and R.V. Churchill in ref.
\cite{Bro03}.}
\label{polos}
\end{figure}

In general, solutions of the Schr\"odinger equation associated to complex
eigenvalues and fulfilling purely outgoing conditions are known as {\it
Gamow-Siegert functions\/}. If $u$ is a function solving $Hu=\epsilon u$,
the appropriate boundary condition may be written
\be
\lim_{x \rightarrow \pm\infty} \left(u'\mp ik u \right)= \lim_{x
\rightarrow \pm\infty} \{ \left(-\beta \mp ik \right) u \} = 0,
\label{condition}
\ee
with $\beta$ defined as the derivative of the logarithm of $u$:
\be
\beta:= -\frac{d}{dx} \ln u.
\label{beta1}
\ee
Now, let us consider a one-dimensional short-range potential $U(x)$,
characterized by a cutoff parameter $\zeta >0$. The general solution of
(\ref{schro2}) can be written in terms of ingoing and outgoing waves:
\be
u_<:= u (x<-\zeta) = I e^{i k x} + L e^{-i k x}, \qquad u_>:= u (x>\zeta)
= N e^{-i k x} + S e^{ikx}
\label{solgral}
\ee
where the coefficients $I,L,N,S,$ depend on the potential parameters and
the incoming energy $k^2 = 2m \epsilon/\hbar^2$ (the kinetic parameter $k$
is in general a complex number $k= k_R+ik_I$), they are usually fixed by
imposing the continuity conditions for $u$ and $du/dx$ at the points
$x=\pm \zeta$. Among these solutions, we are interested in those which are
purely outgoing waves. Thus, the second term in each of the functions
(\ref{solgral}) must dominate over the first one. For such states,
equation (\ref{current}) takes the form:
\be
j_< = - v \vert u_< \vert^2, \qquad j_> = v \vert u_> \vert^2, \qquad v:=
\frac{\hbar}{2m} (k + \overline{k}) = \frac{\hbar k_R}{m}.
\label{current2}
\ee
This last equation introduces the {\it flux velocity\/} $v$. If $\epsilon$
is a real number $\epsilon =E$, then $k$ is either pure imaginary or real
according to $E$ negative or positive. If we assume that the potential
admits negative energies, we get $k_{\pm} = \pm i \sqrt{\vert 2m E/\hbar^2
\vert}$ and (\ref{current2}) vanishes (the flux velocity $v \propto k_R$
is zero outside the interaction zone). Notice that the solutions
$u_E^{(+)}$, connected with $k_+$, are bounded so that they are in
$L^2({\bf R})$. That is, they are the physical solutions $\varphi$
associated with a discrete set of eigenvalues $2m E_n/\hbar^2 =k_{n,+}^2$
solving the continuity equations for $u$ and $du/dx$ at $x=\pm \zeta$. On
the other hand, {\it antibound states\/} $u^{(-)}_E$ increase
exponentially as $\vert x \vert \rightarrow +\infty$. To exhaust the cases
of a real eigenvalue $\epsilon$, let us take now $2m E/\hbar^2
=\kappa^2>0$. The outgoing condition (\ref{condition}) drops the
interference term in the density
\be
\rho (x;t) = \vert N \vert^2 + \vert S \vert^2 + 2 \vert
\overline{N} S \vert \cos (2\kappa x + {\rm Arg} \, S/N), \qquad  
x>\zeta,
\label{density}
\ee
so that the integral of $\rho = \vert S \vert^2$ is not finite neither in
space nor in time (similar expressions hold for $x <-\zeta$). Remark that
flux velocity is not zero outside the interaction zone. Thereby, $E>0$
provides outgoing waves at the cost of a net outflow $j \neq 0$. To get
solutions which are more appropriate for this nontrivial $j$, we shall
consider complex eigenvalues $\epsilon$. Let us write
\be
\epsilon = E - \frac{i}{2} \, \Gamma, \qquad \epsilon_R \propto k_R^2
-k_I^2, \qquad \epsilon_I \propto 2k_R k_I
\label{epsilon}
\ee
where $2m \epsilon/\hbar^2 = (k_R +ik_I)^2$. According to
(\ref{condition}), the boundary condition for $\beta$ reads now
\be
\lim_{x \rightarrow \pm \infty} \{ -\beta \pm (k_I -ik_R) \}=0   
\label{condbeta}
\ee
so that the flux velocity is $v_+ \propto k_R$ for $x > \zeta$ and $v_-
\propto - k_R$ for $x< -\zeta$. Hence, the ``correct'' direction in which
the outgoing waves move is given by $k_R>0$. In this case, the density
\be
\rho(x;t) \equiv \vert u(x,t)\vert^2 =  e^{-\Gamma t/\hbar} \vert
u(x)\vert^2, \qquad \lim_{x\rightarrow \pm \infty} \rho(x;t) \propto
e^{-\Gamma (t -x/v_\pm)/\hbar}
\label{damped}
\ee
can be damped by taking $\Gamma>0$. Thereby, $k_I \neq 0$ and $k_R \neq 0$
have opposite signs. Since $k_R>0$ has been previously fixed, we have $k_I
<0$. Then, purely outgoing, exponentially increasing functions (resonant
states) are defined by points in the fourth quadrant of the complex
$k$-plane. In general, it can be shown that the {\it transmission
amplitude\/} $S$ in (\ref{solgral}) is a meromorphic function of $k$, with
poles restricted to the positive imaginary axis (bound states) and the
lower half-plane (resonances) \cite{Fer08}. Let $k_n$ be a pole of $S$ in
the fourth quadrant of the $k$-plane, then $-\overline{k}_n$ is also a
pole while $\overline{k}_n$ and $-k_n$ are zeros of $S$ (see
Figure \ref{polos}, left). On the other hand, if $S$ is studied as a
function of $\epsilon$, a Riemann surface of $\epsilon^{1/2} =k$ is
obtained by replacing the $\epsilon$-plane with a surface made up of two
sheets $R_0$ and $R_1$, each cut along the positive real axis and with
$R_1$ placed in front of $R_0$ (see e.g. \cite{Bro03}, pp 337). As the
point $\epsilon$ starts from the upper edge of the slit in $R_0$ and
describes a continuous circuit around the origin in the counterclockwise
direction (Figure \ref{polos}, right), the angle increases from $0$ to
$2\pi$. The point then passes from the sheet $R_0$ to the sheet $R_1$,
where the angle increases from $2\pi$ to $4\pi$. The point then passes
back to the sheet $R_0$ and so on. Complex poles of $S(\epsilon)$ always
arise in conjugate pairs (corresponding to $k$ and $-\overline{k}$) while
poles on the negative real axis correspond to either bound or antibound
states.

%%%%%%%%%%%%%%%%%%%%%%%%%%%%%%

%%%%%%%%%%%%%%%%%%%%%%%%%

\begin{figure}[htb]
\centering
\includegraphics[height=.16\textheight]{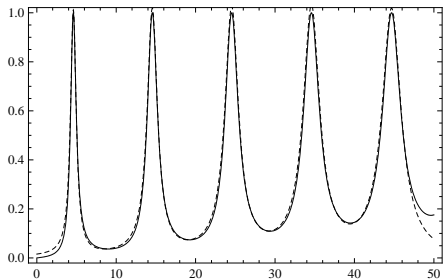}
\caption{\footnotesize Functions $T$ and $\omega_N$ (dotted curve) for a
square well with strength $V_0=992.25$ and weigh $b=20$ for which the FWB
sum matches well the transmission coefficient for the first five
resonances \cite{Fert} (see also \cite{Fer08}).}
\label{encima}
\end{figure}

Observe that density (\ref{damped}) increases exponentially for  
either large $\vert x \vert$ or large negative values of $t$. The
usual interpretation is that the compound ($\varphi_{\epsilon},V$)
represents a decaying system which emitted waves in the remote  
past $t-x/v$. As it is well known, the long lifetime limit
($\Gamma \rightarrow 0$) is useful to avoid some of the
complications connected with the limit $t \rightarrow -\infty$
(see discussions on time asymmetry in \cite{Arn99}). In this
context, one usually imposes the condition:
\be
\frac{\Gamma/2}{\Delta E}<<1.
\label{vecinas}
\ee
Thus, the level width $\Gamma$ must be much smaller than the level spacing
$\Delta E$ in such a way that closer resonances imply narrower widths
(longer lifetimes). In general, the main difficulty is precisely to find
the adequate $E$ and $\Gamma$. However, for one-dimensional stationary
short range potentials, in \cite{Fer08} it has been shown that the
superposition of a denumerable set of FBW distributions (each one centered
at each resonance $E_n, n=1,2,\ldots$) entails an approximation of the
coefficient $T$ such that the larger the number $N$ of close resonances
involved, the higher the precision of the approximation (see
Fig. \ref{encima}):
\be
T \approx \omega_N (\epsilon_R) = \sum_{n=1}^N \omega(\epsilon_R, 
E_n )
\label{taprox}
\ee
with
\be
\omega (\epsilon_R, E) = \frac{(\Gamma/2)^2}{(\epsilon_R -E)^2
+(\Gamma/2)^2}.
\label{bw}
\ee

%%%%%%%%%%%%%%%%%%%%%%%%%%%%%%
%%%%%%%%%%%%%%%%%%%%%%%%%

\noindent
Processes in which the incident wave falls upon a single scatterer are
fundamental in the study of more involved interactions \cite{Nus72}. In
general, for a single target the scattering amplitude is a function of two
variables (e.g. energy and angular momentum). The above model corresponds
to the situation in which one of the variables is held fixed (namely, the
angular momentum). A more realistic three dimensional model is easily
obtained from these results: even functions are dropped while an
infinitely extended, impenetrable wall is added at the negative part of
the straight line \cite{Fer08b}. Such a situation corresponds to $s$-waves
interacting with a single, spherically symmetric, square scatterer (see
e.g. \cite{Fes54}).

%%%%%%%%%%%%%%%
\subsection{Complex Scaling Method}

Some other approaches extend the framework of quantum theory so that
quasi--stationary states can be defined in a precise form. For example,
the complex--scaling method \cite{Agu71,Sim72,Gir03} (see also
\cite{Sud78}) embraces the transformation $H \rightarrow UHU^{-1} =
H_{\theta}$, where $U$ is the complex--scaling operator $U=e^{-\theta
XP/\hbar}$, with $\theta$ a dimensionless parameter and $[X,P]=i\hbar$.
The transformation is achievable by using the Baker-Campbell-Hausdorff
formulae \cite{Mie70}:
\be
e^A B e^{-A} = \left\{e^A, B\right\} = \sum_{n=0}^\infty
\frac1{n!}\left\{A^n,B\right\}
\label{bch}
\ee
with $A$ and $B$ two arbitrary linear operators and
\begin{eqnarray*}
  \left\{A^n, B\right\} &=& \underbrace{\left[A,\left[A,\ldots
\left[A\right.\right.\right.},\left.\left.\left.B\right] 
\ldots\right]\right].
\\
   && \hspace{5mm} n\;\mathrm{times}
\end{eqnarray*}
The identification $A=-\theta XP/\hbar$ and $B=X$ leads to
\be
UXU^{-1}=\sum_{n=0}^\infty \frac1{n!}(i\theta)^n X =e^{i\theta}X, \qquad
UPU^{-1}=\sum_{n=0}^\infty \frac1{n!}(-i\theta)^n P =e^{-i\theta}P,
\label{trans}
\ee
where we have used
\be
\left\{(XP)^n,X\right\} = (-i \hbar)^nX, \qquad \left\{(XP)^n,P\right\} =
(i \hbar)^n P.
\label{parentesis}
\ee
The following calculations are now easy
\be
\begin{array}{l}
UP^2U^{-1} = \left( UPU^{-1}\right) \left(UPU^{-1}\right) =
\left(UPU^{-1}\right)^2=e^{-2i\theta} P^2,\\[2ex]
UV(X)U^{-1} = U\left(\displaystyle \sum_{k=0}^\infty \frac1{k!}V_k
X^k\right)U^{-1} = V\left(e^{i\theta}X \right).
\end{array}
\label{func}
\ee
So that we finally get
\be
UHU^{-1} \equiv H_{\theta}= e^{-2i\theta}P^2 + V(e^{i\theta}X).
\label{complex}
\ee
Remark that in the Schr\"odinger's representation we have
\be
X = x, \quad P= -i\hbar \frac{d}{dx}, \quad U
= e^{i\theta xd/dx} \quad \Rightarrow  \quad Uf(x) =f(xe^{i\theta}).
\label{rep}
\ee

%%%%%%%%%%%%%%%%%%%%%%%%%%%%%%
\begin{figure}[htb]
\centering
\includegraphics[height=.18\textheight]{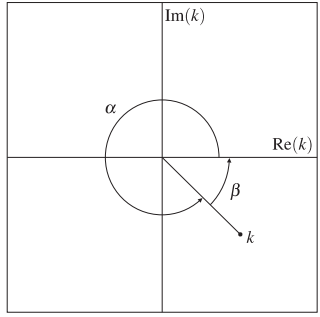}
\caption{\footnotesize Polar form of an arbitrary  point on the complex
$k$-plane.}
\label{kpunto}
\end{figure}
%%%%%%%%%%%%%%%%%%%%%%%%%

\noindent
This transformation converts the description of resonances by
non--integrable Gamow-Siegert functions into one by square integrable
functions. Let $k= \vert k \vert e^{i\alpha} \equiv \vert k \vert
e^{-i\beta}$ be a point on the complex $k$-plane (see
Figure \ref{kpunto}). If $k$ lies on the fourth quadrant then
$0<\beta<\pi/2$ and the related Gamow-Siegert function $u_{\epsilon}$
behaves as follows
\be
u_{\epsilon}(x\rightarrow \pm\infty) \sim e^{\pm i\vert k \vert x \cos
\beta}e^{\pm \vert k \vert x \sin \beta}.
\label{cuadra1}
\ee
That is, $u_{\epsilon}$ diverges for large values of $\vert x \vert$. The
behavior of the complex-scaled function
$\widetilde{u}_{\epsilon}=U(u_{\epsilon})$, on the other hand, reads
\be
\widetilde{u}_{\epsilon}(x\rightarrow \pm\infty) \sim e^{\pm i\vert k
\vert x \cos(\theta -\beta)}e^{\mp \vert k \vert x \sin (\theta -\beta)}.
\label{cuadra2}
\ee
Thereby, $\widetilde{u}_{\epsilon}$ is a bounded function if $\theta -
\beta>0$, i.e., if $\tan \theta > \tan \beta$. The direct calculation
shows that complex-scaling preserves the square-integrability of the
bounded states $\varphi_n$, whenever $0<\theta<\pi/2$. Then one obtains
\be
0<\theta -\beta<\pi/2.
\label{teta}
\ee

%%%%%%%%%%%%%%%%%%%%%%%%%%%%%%
\begin{figure}[ht]
\centering
\includegraphics[height=.18\textheight]{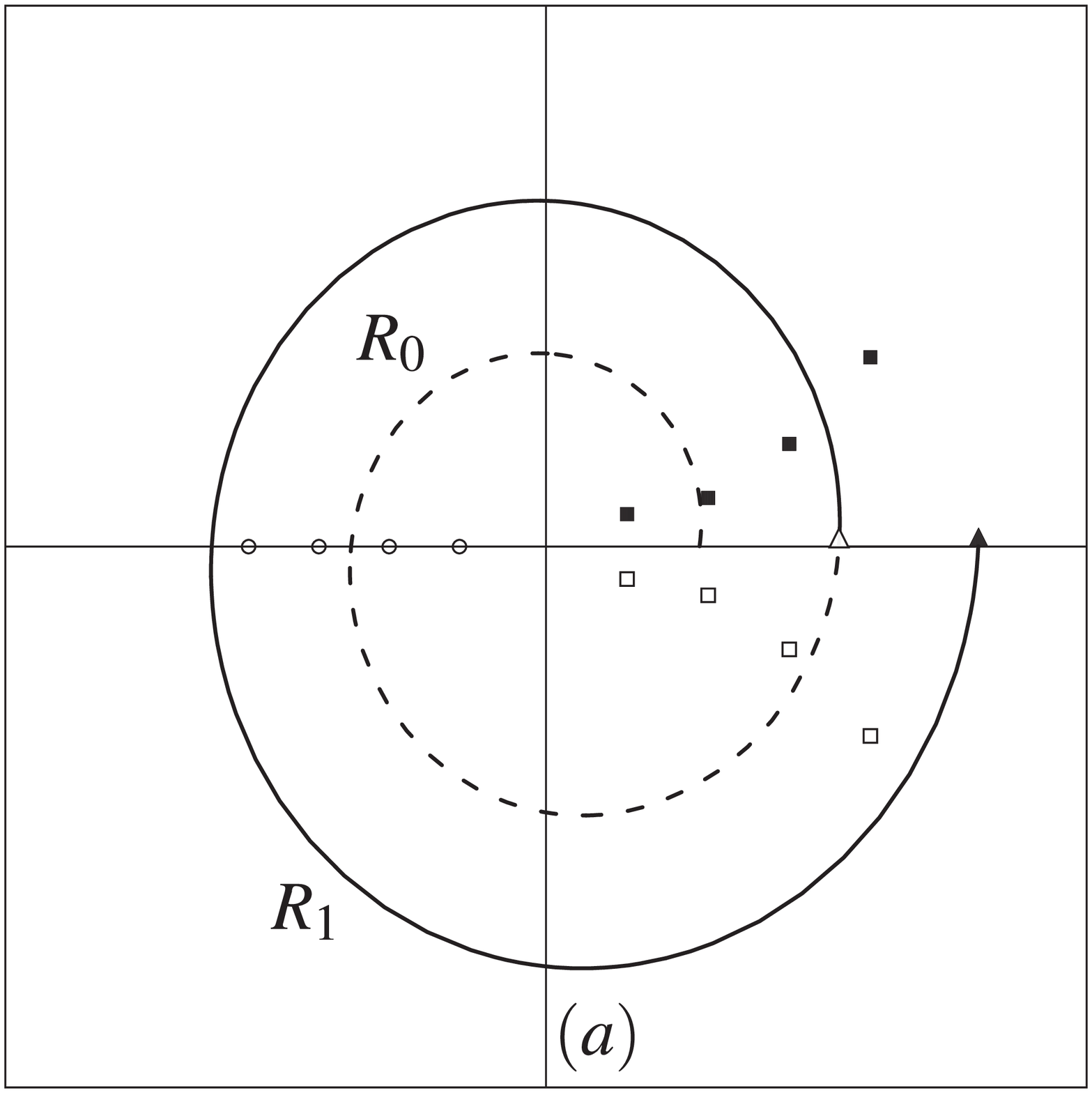} \hskip10mm
\includegraphics[height=.18\textheight]{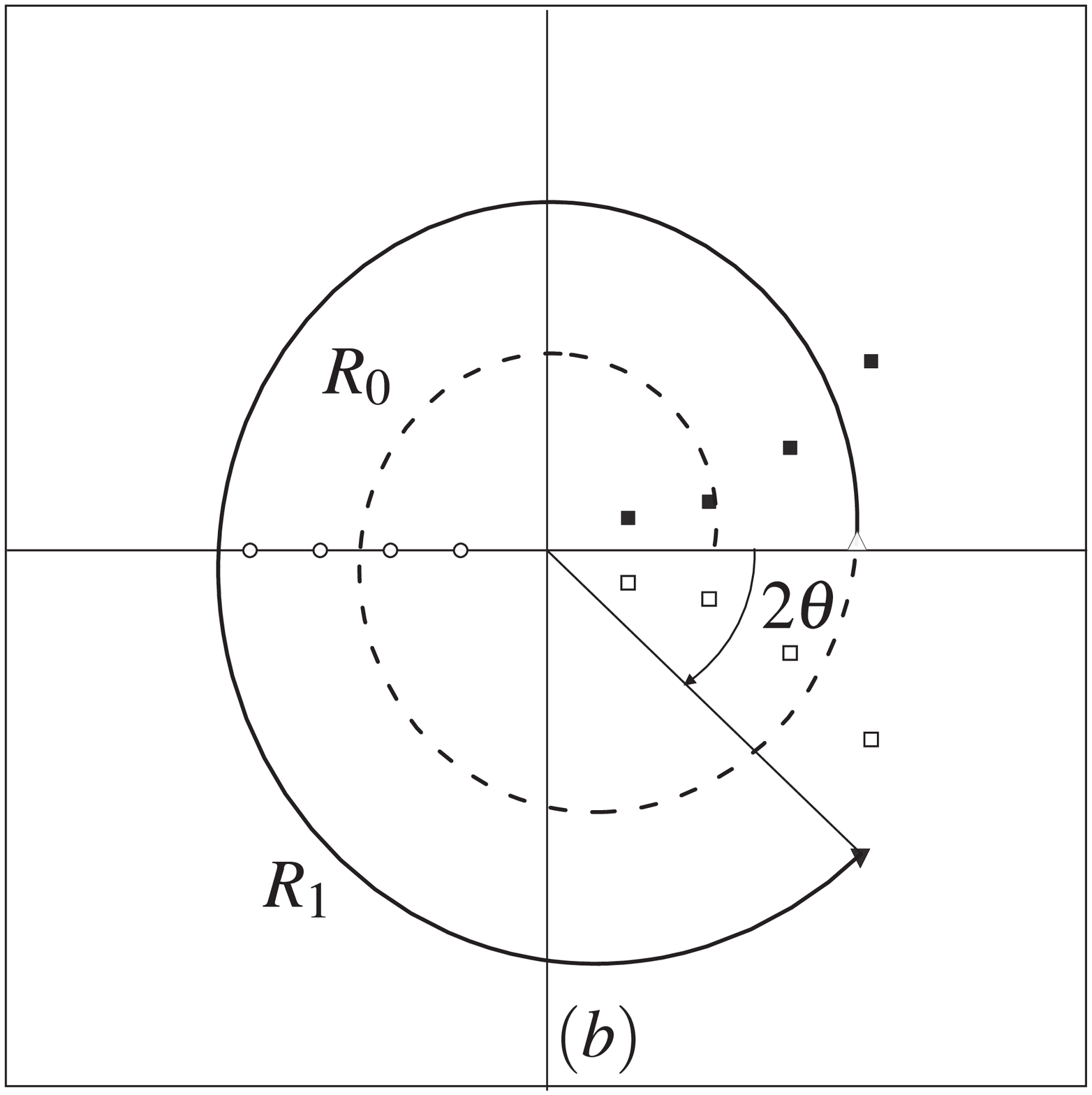}
\caption{\footnotesize {\bf Left.} Schematic representation of the
two-sheet Riemann surface showed in Figure \ref{polos}. Dashed curves and
empty squares lie on the first sheet $R_0$, continuous curves and fulled
squares lie on the second sheet $R_1$ {\bf Right.} The complex rotated
plane `exposing' the resonant poles lying on the first sheet $R_0$.}
\label{cuts1}
\end{figure}
%%%%%%%%%%%%%%%%%%%%%%%%%

\noindent
As regards the complex-scaled scattering states we have
\be
e^{\pm i\vert k \vert x}\rightarrow e^{\pm i \vert k \vert x \cos \theta}
e^{\mp \vert k \vert x\sin \theta}.
\label{scater}
\ee
So that plane waves are transformed into exponential decreasing or
increasing functions for large values of $\vert x \vert$. To preserve the
oscillating form of scattering wave-functions the kinetic parameter $k$
has to be modified. That is, the transformation $k=\vert k \vert
\rightarrow \vert k \vert e^{-i\theta}$ reduces (\ref{scater}) to the
conventional plane-wave form of the scattering states. This
transformation, however, induces a rotation of the positive real axis in
the clockwise direction by the angle $2\theta$: $E \propto k^2 \rightarrow
\vert k \vert^2 e ^{-i2\theta} \propto E e^{-i2\theta}$. That is, the
rotated energy is complex $\epsilon =E_R-i\Gamma/2$ with $E_R = E
\cos(2\theta)$, $\Gamma/2=\sin(2\theta)$. In summary, complex rotation
is such that: 1) Bound state poles remain unchanged under the
transformation 2) Cuts are now rotated downward making an angle of
$2\theta$ with the real axis 3) Resonant poles are `exposed' by the cuts
(see Figure \ref{cuts1}). Another relevant aspect of the method is that it
is possible to construct a resolution to the identity \cite{Ber73}.
Moreover, as the complex eigenvalues are $\theta$--independent, the
resonance phenomenon is just associated with the discrete part of the
complex--scaled Hamiltonian \cite{Moi98} (but see \cite{Sud78}). As a
final remark, let us enphasize that complex-scaling `regularizes' the
divergent Gamow-Siegert functions $u_{\epsilon}$ at the cost of
introducing a non-Hermitian Hamiltonian $H_{\theta}$. From equation
(\ref{complex}) we get
\be
H_{\theta}^{\dagger} = e^{i2\theta} P^2 + V(e^{-i\theta}R) \neq
H_{\theta}.
\label{hcomplex}
\ee
In other words, the `regularized' solutions $\widetilde{u}_{\epsilon}$ are
square-integrable eigenfunctions of a complex potential $V(e^{i\theta}x)$
belonging to the complex eigenvalue $\epsilon$.

%%%%%%%%%%%%%%%%%%%%%%%%%%%
\subsection{Darboux-Gamow Transformations}

In a different survey, complex eigenvalues of Hermitian Hamiltonians have
been used to implement Darboux (supersymmetric) transformations in quantum
mechanics \cite{Can98,Fer03,Ros03,Ros07,Fer07,Sam06a} (see also the
discussion on `atypical models' in \cite{Mie04}). The transformed
Hamiltonians include non-Hermitian ones, for which the point spectrum
sometimes has a single complex eigenvalue
\cite{Fer03,Ros03,Ros07,Fer08,Fer08b}. This last result, combined with
appropriate squeezing operators \cite{Fer00}, could be in connection with
the complex-scaling technique. In general, supersymmetric transformations
constitute a powerful tool in quantum mechanics \cite{Mie04}. However, as
far as we know, until the recent results reported in
\cite{Fer08,Fer08b,Ros07,Fer07} the connection between
supersymmetric transformations and resonant states has been missing. In
this context and to throw further light on the complex function $\beta$ we
may note that (\ref{beta1}) transforms the Schr\"odinger equation
(\ref{schro2}) into a Riccati one
\be
-\beta'+\beta^2+\epsilon=V
\label{riccati}
\ee
where we have omitted the units. Remark that (\ref{riccati}) is not
invariant under a change in the sign of the function $\beta$:
\be
\beta' + \beta^2 + \epsilon= V + 2 \beta'.
\label{nricatti}
\ee
These last equations define a Darboux transformation $\widetilde{V} \equiv
\widetilde{V}(x, \epsilon) = V(x) +2\beta'(x)$ of the initial potential
$V$. This transformation necessarily produces a complex function if $u$ in
equation (\ref{beta1}) is a Gamow-Siegert function $u_{\epsilon}$. That
is, a {\it Darboux-Gamow deformation\/} is defined as follows
\cite{Fer08}:
\be
\widetilde{V}=V+2\beta' \equiv V - 2\frac{d^2}{dx^2} \ln
u_{\epsilon}.
\label{Vtilde}
\ee
The main point here is that the purely outgoing condition
(\ref{condition}) leads to $\beta' \rightarrow 0$ so that $\widetilde V
\rightarrow V$, in the limit $\vert x \vert \rightarrow +\infty$. In
general, according with the excitation level of the transformation
function $u_{\epsilon}$, the real $\widetilde V_R$ and imaginary
$\widetilde V_I$ parts of $\widetilde V$ show a series of maxima and
minima. Thus, the new potential behaves as an optical device emitting and
absorbing probability flux at the same time, since the function $I_I(x)$
shows multiple changes of sign \cite{Fer08,Fer08b,Ros07,Fer07}. On the
other hand, the solutions $y \equiv y(x,\epsilon,{\cal E})$ of the
non-Hermitian Schr\"odinger equation
\be
-y''+\widetilde{V}y={\cal E} y
\label{Schrod2}
\ee
are easily obtained
\be
y \propto \frac{{\rm
W}(u_{\epsilon},\psi)}{u_{\epsilon}},
\label{1ssol}
\ee
where ${\rm W}(*,*)$ stands for the Wronskian of the involved functions
and $\psi$ is eigen-solution of (\ref{Schrod2}) with eigenvalue ${\cal E}$.
It is easy to show that scattering waves and their Darboux-Gamow
deformations share similar transmission probabilities \cite{Fer08}. Now,
let us suppose that Hamiltonian $H$ includes a point spectrum $\sigma_d(H)
\subset {\rm Sp} (H)$. If $\psi_n$ is a (square-integrable) eigenfunction
with eigenvalue ${\cal E}_n$, then its Darboux-Gamow deformation
(\ref{1ssol}) is bounded:
\be
\lim_{x \rightarrow \pm \infty} y_n = \mp (\sqrt{{\cal E}_n} +
ik)(\lim_{x \rightarrow \pm \infty} \psi_n ).
\label{bound}
\ee
Thereby, $y_n$ is a normalizable eigenfunction of $\widetilde H$
with eigenvalue ${\cal E}_n$. However, as $\epsilon$ is complex,  
although the new functions $\{ y_n \}$ may be normalizable, they
will not form an orthogonal set \cite{Ros03} (see also
\cite{Ram03} and the `puzzles' with self orthogonal states
\cite{Sok06}). There is still another bounded solution to be
considered. Function $y_{\epsilon} \propto
\varphi^{-1}_{\epsilon}$ fulfills equation (\ref{Schrod2}) for the
complex eigenvalue $\epsilon$. Since $\lim_{x \rightarrow \, \pm  
\infty} \vert y_{\epsilon}\vert^2 = e^{\pm 2 k_I x}$ and $k_I<0$,
we have another normalizable function to be added to the set
$\{y_n\}$.

In summary, one is able to construct non-Hermitian Hamiltonians
$\widetilde{H}$ for which the point spectrum is also
$\sigma_d(H)$, extended by a single complex eigenvalue ${\rm Sp}
(\widetilde{H}) = {\rm Sp} (H) \cup \{\epsilon\}$. As we can see, the
results of the Darboux-Gamow deformations are quite similar to those
obtained by means of the complex-scaling method. This relationship
deserves a detailed discussion which will be given elsewhere.

%%%%%%%%%%%%%%%%%%%%%%%%%%%%%
\section{Conclusions}

We have studied the concept of {\it resonance\/} as it is understood in
classical mechanics by analyzing the motion of a forced oscillator with
damping. The resonance phenomenon occurs for steady state oscillations
when the driving force oscillates at an angular frequency equal to the
natural frequency $w_0$ of the oscillator. Then the amplitude of the
oscillation is maximum and $w_0$ is called the resonance frequency. The
spectral energy distribution corresponds to a Fock-Breit-Wigner (FBW)
function, centered at $w_0$ and having a line breadth equal to the damping
constant $\gamma$. Resonance is present even in the absence of
external forces (transient oscillations). In such case the energy
decreases exponentially with the time so that the damping constant
$\gamma$ is a measure of the lifetime of the oscillation $\tau =
1/\gamma$. Similar phenomena occur for the electromagnetic radiation. In
this context, the effects of radiative radiation can be approximated by
considering the radiative reaction force $\vec F_{\rm rad}$ as a friction
force which damps the oscillations of the electric field. Thus, the
concept of resonance studied in classical mechanics is easily extended to
the Maxwell's electromagnetic theory. The model also applies in vibrating
elastic bodies, provided that the displacement is now a measure of the
degree of excitation of the appropriate vibrational mode of the sample.
{\it Acoustic resonances\/} are then obtained when the elastic bodies
vibrate in such a way that standing waves are set up (Some interesting
papers dealing with diverse kinds of resonances in metals can be consulted
in \cite{Now64}). Since the profile of atomic phenomena involving a high
number of quanta of excitation can be analyzed in the context of classical
mechanics, cyclotron and electron spin resonances can be studied, in a
first approach, in terms of the above model (see, e.g. the paper by A.S.
Nowick in \cite{Now64}, pp 1-44, and references quoted therein). The
quantum approach to the problem of spinning charged particles showing
magnetic resonance is discussed in conventional books on quantum mechanics
like the one of Cohen-Tannoudji et. al. \cite{Coh77}.

We have also shown that the resonance phenomenon occurs in quantum
decaying systems. According with the Fock's approach, the corresponding
law of decay depends only on the energy distribution function $\omega(E)$
which is meromorphic and acquires the form of a FBW, bell-shaped curve.
The introduction of $\epsilon = E-i\Gamma/2$, a complex eigenvalue of the
energy, is then required in analyzing the resonances which, in turn, are
identified with the decaying states of the system. The inverse of the
lifetime is then in correspondence with $\Gamma/2$. In a simple model, the
related exponential decreasing probability can be justified by introducing
a complex potential $U = U_R +U_I$, the imaginary part of which is the
constant $-\Gamma/2$. Thus, $U_I$ works like a sink of probability waves.
The situation resembles the absorption of electromagnetic waves by a
medium with complex refractive index so that $U= U_R +U_I$ is called an
optical potential in nuclear physics. The treatment of complex energies in
quantum mechanics includes non-square integrable Gamow-Siegert functions
which are outside of the Hilbert spaces. In this sense, the
complex-scaling method is useful to `regularize' the problem by
complex-rotating positions $x$ and wavenumbers $k$. As a consequence,
bounded and scattering states maintain without changes after the rotation
while the Gamow-Siegert functions become square-integrable. Another
important aspect of the method is that the positive real axis of the
complex energy plane is clockwise rotated by an angle $2\theta$, so that
resonant energies are exposed by the cuts of the corresponding Riemann
surface. The method, however, also produces complex potentials. That is,
the Gamow-Siegert functions are square-integrable solutions of a
non-Hermitian Hamiltonian belonging to complex eigenvalues. Similar
results are obtained by deforming the initial potential in terms of a
Darboux transformation defined by a Gamow-Siegert function. In this sense,
both of the above approaches could be applied in the studying of quantum
resonances. A detailed analysis of the connection between the
complex-scaling method and the Darboux-Gamow transformation is in
progress.

%%%%%%%%%%%%%%%%%%%%%%%%%%%%%%%%%%%%%%%%%%%%%%%%
%% BACKMATTER
%%%%%%%%%%%%%%%%%%%%%%%%%%%%%%%%%%%%%%%%%%%%%%%%

\section*{Acknowledgments}

ORO would like to thank to the organizers for the kind invitation to such
interesting Summer School. Special thanks to Luis Manuel Monta\~no and
Gabino Torres. SCyC thanks the members of Physics Department, Cinvestav,
for kind hospitality The authors wish to thank Mauricio Carbajal for his
interest in reading the manuscript. The support of CONACyT project
24233-50766-F and IPN grants COFAA and SIP (IPN, Mexico) is acknowledged.

%%%%%%%%%%%%%%%%%%%%%%%%%%%%%%%%%%%%%%%%%%%

\end{document}